\documentclass[sigconf]{acmart}
\AtBeginDocument{%
  }

\copyrightyear{2026}
\acmYear{2026}
\setcopyright{cc}
\setcctype{by}
\acmConference[CHI '26]{Proceedings of the 2026 CHI Conference on Human Factors in Computing Systems}{April 13--17, 2026}{Barcelona, Spain}
\acmBooktitle{Proceedings of the 2026 CHI Conference on Human Factors in Computing Systems (CHI '26), April 13--17, 2026, Barcelona, Spain}
\acmPrice{}
\acmDOI{10.1145/3772318.3791774}
\acmISBN{979-8-4007-2278-3/2026/04}

\usepackage{wrapfig}
\usepackage{float}
\usepackage{array}
\usepackage{longtable}
\usepackage{booktabs}

\begin{document}

\title{Beyond the Checkbox: Strengthening DSA Compliance Through Social Media Algorithmic Auditing}

\author{Sara Solarova}
\affiliation{
  \institution{Kempelen Institute of Intelligent Technologies}
  \city{Bratislava}
  \country{Slovakia}}
\email{sara.solarova@kinit.sk}
\orcid{0000-0003-2800-3572}

\author{Matúš Mesarčík}
\affiliation{
  \institution{Kempelen Institute of Intelligent Technologies}
  \city{Bratislava}
  \country{Slovakia}
}
\affiliation{
  \institution{Comenius University Bratislava}
  \city{Bratislava}
  \country{Slovakia}
}
\email{matus.mesarcik@kinit.sk}
\orcid{0000-0003-3311-5333}

\author{Branislav Pecher}
\affiliation{
  \institution{Kempelen Institute of Intelligent Technologies}
  \city{Bratislava}
  \country{Slovakia}}
\email{branislav.pecher@kinit.sk}
\orcid{0000-0003-0344-8620}

\author{Ivan Srba}
\affiliation{
  \institution{Kempelen Institute of Intelligent Technologies}
  \city{Bratislava}
  \country{Slovakia}}
\email{ivan.srba@kinit.sk}
\orcid{0000-0003-3511-5337}

\renewcommand{\shortauthors}{Solarova et al.}

\begin{abstract}
    Algorithms of online platforms are required under the Digital Services Act (DSA) to comply with specific obligations concerning algorithmic transparency, user protection and privacy. To verify compliance with these requirements, DSA mandates platforms to undergo independent audits. Little is known about current auditing practices and their effectiveness in ensuring such compliance. To this end, we bridge regulatory and technical perspectives by critically examining selected audit reports across three critical algorithmic-related provisions: restrictions on profiling minors, transparency in recommender systems, and limitations on targeted advertising using sensitive data. Our analysis shows significant inconsistencies in methodologies and lack of technical depth when evaluating AI-powered systems. To enhance the depth, scale, and independence of compliance assessments, we propose to employ algorithmic auditing – a process of behavioural assessment of AI algorithms by means of simulating user behaviour, observing algorithm responses and analysing them for audited phenomena.
\end{abstract}





\maketitle

\section{Introduction}
In 2021, Facebook whistleblower Frances Haugen released internal documents revealing that Meta’s recommender system actively amplified divisive and harmful content, prioritizing engagement over user well-being \cite{Mac2021FacebookClamps}. Internal documents showed that inflammatory posts received greater algorithmic promotion, even when they contained misinformation or extremist rhetoric. Similar concerns have been raised about TikTok’s \textit{For You Page}, where algorithmic recommendations have been shown to push harmful content including eating disorder-related videos to vulnerable users within minutes of account creation \cite{Rawlinson2023TikTokChildren}. In 2024, TikTok’s algorithms appear to have significantly affected electoral dynamics in Romania by enabling the rapid amplification of fringe and extremist messaging \cite{social_media_dynamics}. In particular, its recommendation system, supported by networks of inauthentic accounts, helped boost the visibility of a once-marginal candidate.

Despite public assurances of content moderation and algorithmic oversight, these revelations demonstrated a fundamental gap between platform policies and actual algorithmic behaviour. In addition, they exemplify how recommender systems and search engines, operating as black-box AI models, shape user experiences and influence public discourse in ways that remain largely opaque to regulators and the public. These cases underscore the urgent need for robust oversight mechanisms.

The European Union’s \textit{Digital Services Act} (DSA) legislation \cite{EU2022DSA} seeks to address these challenges by imposing transparency and accountability obligations on digital service providers including \textit{Very Large Online Platforms} (VLOPs) and \textit{Very Large Online Search Engines} (VLOSEs). Among its key provisions, the DSA mandates independent annual audits for VLOPs and VLOSEs to assess compliance with imposed obligations focused, besides others, on algorithmic transparency or user privacy and security.

The actual implementation of auditing practices employed by external auditors to assess algorithmic compliance, however, remains an underexplored area \cite{Sekwenz2025DoingAuditsRight}. This knowledge gap limits our ability to assess the adequacy of current regulatory frameworks and identify potential improvements needed for effective algorithmic governance.

To address this shortcoming, this analysis examines the first wave of audit reports published in Q4 2024 from YouTube (Google Ireland Limited), Facebook and Instagram (Meta Platforms Ireland Limited), and TikTok (TikTok Technology Limited) to identify systematic challenges in how auditors assess compliance with critical algorithmic-related provisions regarding recommender system transparency (Articles 27(3) and 38), minors protection (Article 28(2)), and advertising restrictions (Article 26(3)).

The central observation of this analysis is that current \textit{traditional audit methodologies}, rooted in compliance auditing traditions developed for financial reporting and IT controls, were designed for relatively static systems and therefore face unprecedented challenges when applied to modern AI-based systems that  interact dynamically with human users and often exceed the scope of conventional audit approaches. By examining the specific audit reports conducted across platforms, we identify two critical issues: (1) significant inconsistencies in audit approaches and outcomes, and (2) the absence of advanced methodologies capable of effectively assessing complex algorithmic behaviours. These findings reveal why current, conventional approaches may be structurally insufficient for achieving the accountability that DSA regulation envisions.

The main contributions of this article are as follows:
\begin{enumerate}
    \item We conduct a so-far-missing systematic examination of actual DSA audit reports to demonstrate significant inconsistencies in methodologies and substantial variations in technical depth when evaluating AI-powered systems across platforms. This study provides systematic empirical analysis of how auditors operationalise technically complex compliance requirements in practice. Thereby, we bridge regulatory and technical perspectives, providing both computer science and legal experts with insights into the methodological challenges of auditing algorithmic systems and guidance for developing more robust audit approaches. Additionally, this paper extends existing human-computer interaction (HCI) research by translating its methodological principles into the regulatory context of the DSA, highlighting where current legal audits diverge from established best practices.
    
    \item To address these fundamental limitations, we propose algorithmic auditing\footnote{We use ``algorithmic auditing'' to denote a method of systematic querying an algorithm with inputs and observing the corresponding outputs (e.g., through controlled probing and user simulation). Some researchers use different terms of ``algorithm audit'' or ``algorithm auditing''  \cite{Metaxa2021AuditingAlgorithms, Lam2023SociotechnicalAudits, Sandvig2014Audits}, 
    while referring to the same concept. Our selection of ``algorithmic auditing'' follows its adoption in regulatory and legal contexts \cite{GoodmanTrehu2023AlgorithmicAuditing}, which this work is positioned in.} as a process of behavioural assessment through user behaviour simulation and systematic response analysis, specifically designed to overcome the observability constraints that traditional audit methods encounter when evaluating dynamic AI systems.
\end{enumerate}

This paper proceeds as follows. We begin by reviewing related work on platform auditing, examining the sociotechnical challenges of algorithmic oversight and pointing out potential limitations in traditional audit methodologies. We then outline our methodology for analysing publicly available DSA audit reports from major platforms. Our results provide systematic analysis of audit reports across three critical compliance areas: recommender system transparency and user control options (Articles 27 and 38), protection of minors from profiling-based advertising (Article 28), and restrictions on advertisements using sensitive personal data (Article 26). Through detailed examination of how different auditors approached identical regulatory requirements across YouTube, Facebook, Instagram, and TikTok, we identify significant methodological variations and assessment limitations. We discuss these findings in light of the structural challenges that traditional audit approaches face when applied to dynamic algorithmic systems. To address identified challenges, we propose emerging algorithmic auditing methods as a potential solution. To demonstrate how such an approach can considerably strengthen algorithmic oversight, we provide an illustrative use case for auditing a protection of minors from profiling-based advertising (Article 28). The paper concludes by examining the broader implications for DSA enforcement and the need for new approaches to effectively implement the DSA regulation.

\section{Background and Related Work}

\subsection{DSA Obligations and Audits Verifying Regulation Compliance}
Digital Service Act (DSA) \cite{EU2022DSA} introduces a far-reaching regulatory regime designed to promote greater safety, transparency, and accountability in the EU’s digital sphere. Its underlying rationale is to respond to the profound societal and economic influence of online platforms, with particular attention to the biggest players -- VLOPs and VLOSEs. The DSA regulation seeks to safeguard fundamental rights, provide rules for liability for third party content and moderation practices, imposes clear responsibilities for recommender systems, curbs manipulative design practices, and establishes mechanisms to mitigate systemic risks.

\begin{figure}[t]
  \centering
  \includegraphics[width=0.65\linewidth]{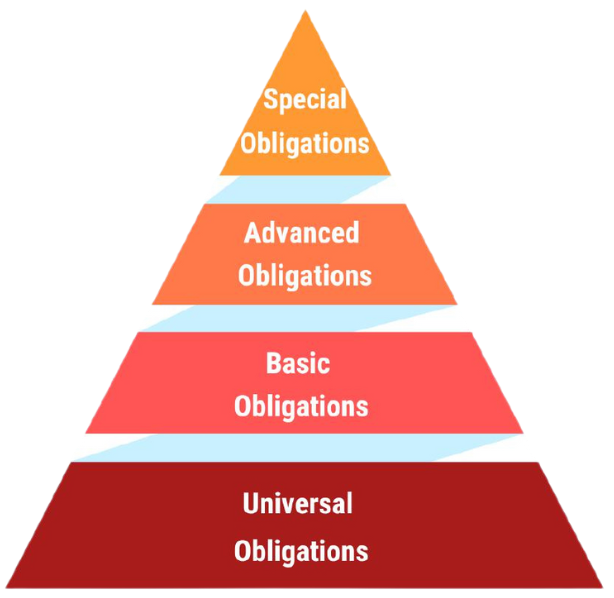}
  \caption{A DSA tiered system of due diligence obligations (inspired by \cite{HusovecLaguna2022DigitalServicesAct}). It is structured as a regulatory pyramid: 1) at its base lie the universal and basic obligations that apply to all intermediary services, complemented by advanced obligations for hosting providers and online platforms; 2) at the top of this hierarchy stand the special obligations tailored to VLOPs and VLOSEs, which are specifically crafted to address systemic risks arising from their scale and societal reach.}
  \Description{A tiered system of due diligence obligations for digital services is structured as a regulatory pyramid: 1) at its base lie the universal and basic obligations that apply to all intermediary services, complemented by advanced obligations for hosting providers and online platforms; 2) at the top of this hierarchy stand the special obligations tailored to VLOPs (and VLOSEs), which are specifically crafted to address systemic risks arising from their scale and societal reach.}
  \label{fig:dsa-tiers}
\end{figure}

The DSA adopts a tiered system of \textit{due diligence obligations} for digital services, structured as a regulatory pyramid (Figure \ref{fig:dsa-tiers}; see \cite{deMiguelAsensio2022DueDiligence, HusovecLaguna2022DigitalServicesAct} for more details). At its base lie the universal obligations that apply to all intermediary services. They are further complemented by basic obligations for hosting providers and advanced obligations for online platforms. The top of this hierarchy comprises of the special obligations tailored to VLOPs and VLOSEs, which are specifically crafted to address \textit{systemic risks} arising from their scale and societal reach. This layered approach creates an integrated framework of responsibilities that grows progressively stricter in line with the role, size and impact of the service provider.

Apart from legal obligations mentioned above, the DSA requires VLOPs and VLOSEs to undergo \textit{annual external audits}. These audits are intended to provide an independent assessment of how these entities comply with their legal responsibilities (DSA, Article 37), particularly regarding risk assessment and the mitigation of systemic risks (DSA, Article 33 and 34). An audit, in general terms, is a structured, objective, and documented process for gathering and evaluating evidence to determine whether specific criteria are met. External audits may be classified as either second-party, when conducted by stakeholders such as business partners or customers, or third-party, when undertaken by independent certification bodies or public authorities. Article 37 DSA expressly requires that VLOPs and VLOSEs be regularly examined by qualified and independent \textit{auditors}. These evaluations enhance accountability and transparency by producing detailed \textit{audit reports} containing recommendations for remedial measures. Such reports are transmitted to the European Commission and may prompt enforcement actions where deficiencies are identified.

The main purpose of this oversight is to verify compliance with the regulation's obligations, with emphasis on the handling of systemic risks, including the spread of illegal content, protection of minors, and safeguarding of public interests. At its core is the duty to manage systemic risks, which obliges VLOPs and VLOSEs to identify, assess, and reduce harms that may arise from their operations (DSA, Article 34). These risks encompass not only illegal content and infringements of fundamental rights, but also threats to democratic discourse, privacy, security, public order, health, minors, and overall user well-being. Auditors must therefore determine whether platforms have put in place effective, evidence-based mitigation strategies, whether these strategies are implemented in practice, and whether their reporting is transparent, consistent, and methodologically sound.

The special obligations linked to systemic risk management are embedded in a wider compliance architecture that extends from universal and basic obligations -- such as mechanisms for content moderation, user notifications, and regulatory cooperation -- to advanced requirements concerning trusted flaggers, recommender system transparency, and protections against manipulative design. Audits serve to test the integrity of this interconnected framework, ensuring that each element works together to reduce risks and protect users’ rights. In addition, auditors are required to assess the extent to which VLOPs and VLOSEs comply with voluntary commitments, including participation in Codes of Conduct (DSA, Article 45) and Crisis Protocols (DSA, Article 48).

The specific procedures for carrying out audits are detailed in the Commission’s \textit{delegated regulation} on audit rules \cite{EU2024_436}. This implementing act sets out the methodologies, criteria, and procedural steps to be followed in assessing compliance with the DSA, particularly in areas such as risk management, transparency, and content moderation. It further defines the independence and professional qualifications auditors must meet, delineates the audit’s scope, and introduces a structured framework for planning, conducting, and reporting audits. The overarching aim of these rules is to guarantee that audits are rigorous, reliable, and harmonised across the Union, thereby holding VLOPs and VLOSEs effectively accountable for the systemic risks their services generate. 

For the purpose of auditing and compliance assessing, DSA does not provide specific thresholds of \textit{evidentiary standard}. This is evaluated on case-by-case basis by the European Commission tasked with oversight obligations over VLOPs and VLOSEs. Such open-textured approach mainly relying on proportionality, professional judgment, and the state-of-the-art approaches, results in significant discretion in how audits are conducted. Delegated regulation provides only two general requirements in terms of evidentiary standard \cite{EU2024_436}. Firstly, evidence shall be relevant and sufficient to substantiate risk analysis and provide basis for reasonable conclusions. Secondly, the evidence shall be reliable. As a result, evidentiary expectations remain variable and uncertain, creating significant divergence in how auditors may interpret and apply the DSA’s obligations. At the same time, we acknowledge that over-standardization risks creating rigid frameworks unable to adapt to platform diversity and evolving algorithmic practices. This creates a tension that requires ongoing calibration between comparability and contextual responsiveness.

\subsection{Existing Audit Methodologies}
\textbf{Traditional Audits}. The professional service firms who currently conduct DSA audits employ a paradigm of \textit{traditional (conventional)} auditing. It is a verification practice in which auditors assess whether an organisation adheres to regulatory requirements by examining whether appropriate controls, policies, and procedures exist and function as documented. The underlying premise is that compliance can be established by confirming the presence and proper documentation of controls - an approach developed for relatively static systems such as financial reporting and IT infrastructure, where documented procedures reliably predict operational behaviour.

More specifically, the first wave of the DSA audits were conducted in accordance with ISAE 3000 (International Standard on Assurance Engagements Other Than Audits or Reviews of Historical Financial Information) \cite{ISAE3000}, as explicitly stated in the audit reports themselves. This standard prescribes a repertoire of evidence-gathering techniques inherited from financial auditing practice (ISA 500) \cite{ISA500}:
\begin{itemize}
    \item Inspection: examining records, documents, policies, samples of code, or system configurations to verify their existence and content.
    \item Observation: watching a process or procedure being performed by others at a specific moment.
    \item Inquiry: seeking information from knowledgeable persons inside or outside the entity, typically through management interviews. 
    \item Confirmation: obtaining corroborating evidence directly from third parties.
    \item Re-performance: independently executing a procedure to verify it functions as documented. 
\end{itemize}

\textbf{The Limits of Traditional Audit Methodologies}. These techniques share a critical characteristic: they are designed as point-in-time assessments to verify that controls exist and operate correctly at the moment of examination. This design reflects the origins of ISAE 3000 in financial auditing, where the underlying systems being assessed (ledgers, reporting controls, approval workflows) are relatively static: a control that functions correctly today will, barring deliberate change, function correctly tomorrow.

Emerging research works, providing the empirical evidence from early DSA audits, reinforce these concerns. Previous analyses reveal significant inconsistencies across platforms because key terms such as “plain and intelligible language” and “most significant” recommender system criteria remain undefined, leaving platforms and auditors to develop their own interpretations and benchmarks \cite{chapman2025ambiguity}. Seemingly small methodological choices, including account parameters, session duration, or segmentation strategies, can meaningfully alter conclusions about algorithmic behaviour \cite{bouchaud2024auditing} exposing sensitivities that conventional IT audit techniques were never designed to handle. Additional work shows that platform-provided explanations may be unreliable and misleading. TikTok, for example, applied the explanation “You commented on similar videos” to 34 percent of videos shown to accounts that had never left any comments on the platform \cite{mousavi2024auditing}. This creates circular validation problems where auditors assess compliance using potentially inaccurate information provided by the systems they are evaluating.

Structural risks also persist. Even though DSA Article 37 establishes audits of VLOPs and VLOSE to be independent, these mechanisms could still be subject to an \textit{audit capture} phenomenon. Here, they introduce the risk of auditors potentially being financially or structurally dependent on the platforms they audit, which could lead to biased results \cite{laux2021taming}. Though some scholars call for stricter safeguards to ensure auditor independence \cite{laux2021taming}, alternatives like rotating audit firms, state-funded independent auditing bodies, or adversarial auditing systems remain overlooked. This gap only further exacerbates the limitations of current auditing practices, which fail to address the fundamental risks of manipulation and compromised accountability.

Further challenges follow from the scale and complexity of platform operations: rather than a single algorithm, platforms rely on interconnected systems \cite{messmer2023auditing} making sampling indispensable yet difficult. Effective sampling must navigate geographic, linguistic, temporal, and structural variation, but analyses of actual audit reports show substantial divergence in sampling design and transparency, with some firms relying heavily on self-reported data \cite{sekwenz2025doing}.

These gaps illustrate how currently dominant traditional auditing methodologies treat platforms as static systems and demonstrate \textit{the need for incorporating additional approaches that allow independent oversight and fully account for inevitable sociotechnical complexity.}

\textbf{Sociotechnical Challenge}.
The overarching consensus among social scientists and ethicists is that platform governance cannot be understood through purely technical lenses. Sociotechnical transparency as a concept explains that algorithmic transparency is not merely about code disclosure but equally about comprehending how technology interacts with social, political, and economic systems to shape online discourse and behaviour \cite{leerssen2020soap}. This perspective challenges audit methodologies that focus on technical compliance verification, suggesting that meaningful oversight must capture the complex feedback loops between algorithmic processes and social behaviours.

The risk-scenario-based approach operationalises the sociotechnical perspective through a four-step methodology (planning, defining scenarios, developing measurements, evaluating), capturing interactions between technical systems and social behaviours. Although the approach incorporates multi-stakeholder engagement, its operationalisation still relies on conventional techniques mentioned earlier, such as code reviews, document analysis, interface testing, which may prove inadequate for assessing complex sociotechnical phenomena \cite{messmer2023auditing}.

Furthermore, scholars examining the DSA have placed particular emphasis on the role of researchers~\cite{Husovec2023DataAccess, Liesenfeld2025IndependentResearch, Jaursch2023ResearcherAccess, Turillazzi2023DSAEthics}, both in the framework of the regulation generally~\cite{FrosioObafemi2025AugmentedAccountability} and in more specific contexts including intersections with the Act on Digital Markets, political economy of algorithmic auditing or risk-based scenarios~\cite{Laux2021TamingTheFew, Terzis2024AlgorithmicAudits, messmer2023auditing, Sekwenz2025DoingAuditsRight}. Academic debate has especially highlighted the importance of AI auditing environments~\cite{Hartmann2025AIecosystem} and HCI-related practices, such as \textit{algorithmic audits}~\cite{GoodmanTrehu2023AlgorithmicAuditing,Panigutti2025DSAReview}.

\textbf{Emerging Approaches Rooted in Human-Computer Interaction Research}.
HCI research has developed a diverse set of methods for examining algorithmic systems as sociotechnical constructs, emphasising that technical outputs cannot be separated from user behaviour, platform design, and organisational context. \citet{Metaxa2021AuditingAlgorithms} outlines core algorithmic audit methodologies including controlled probing, automated browser collection, and ecological user-side data gathering and subsequently highlights recurring challenges such as personalisation noise, baselining, and limited interpretability. Sociotechnical audits proposed by \citet{Lam2023SociotechnicalAudits} extend this perspective through large-scale user studies that reveal how algorithmic harms emerge through patterns of exposure and interaction, demonstrating that many risks cannot be detected through input-output auditing alone. Further research investigates how everyday users, rather than experts, can surface harmful algorithmic behaviours, suggesting user-driven auditing as a complement to traditional audits \cite{devos} and the importance of end users auditing algorithms themselves \cite{lamenduser}. 
At the same time, research conducted by \citet{Deng2023UserEngagedAuditing} shows that industry practitioners increasingly experiment with user-engaged auditing, emphasizing uncovering blindspots and subjective experiences of harm, while facing organisational and legal barriers that prevent systematic application. More recent regulation-related work, such as \citet{Panigutti2025DSAReview} focusing on the review of the DSA, surveys audit study designs and identifies structural limitations, most specifically the continued reliance on indirect testing and the absence of transparent access to platform internal infrastructures. Altogether, HCI methods offer techniques that address gaps in traditional auditing but they remain fragmented, experimental, and insufficiently integrated into formal compliance processes.

We can conclude that despite the growing attention, the \textit{practical implementation of auditing methodologies in DSA compliance assessment remains largely underexplored}. Published audit reports provide a unique empirical window into these practices, yet systematic analysis of such documents is still largely absent from the literature.

\section{Methodology}
This study examines DSA compliance audit methodologies through \textit{qualitative analysis} of publicly available independent audit reports (DSA Article 37) from the selected largest designated VLOPs.

We adopted a qualitative \textit{document analysis} approach \cite{bowen2009document}, treating the published audit reports as primary documentary evidence of how auditors operationalise DSA compliance requirements in practice. Document analysis is particularly suited to this inquiry as it enables systematic examination of official institutional records that would otherwise remain opaque to external scrutiny \cite{bowen2009document, prior2003using}. Following Bowen's recommendation \cite{bowen2009document}, our analytical strategy integrates directed content analysis \cite{hsieh2005three} with thematic analysis procedures \cite{braun2006thematic}. The DSA's specific articles (26, 27, 28, and 38) provided a deductive framework for initial categorisation, while patterns regarding methodological inconsistencies and assessment limitations emerged inductively through iterative engagement with the data.
Our design is explicitly comparative in a way that by examining how different auditors approached identical regulatory requirements across audited platforms, we can identify both convergent practices and methodological variations that might otherwise remain invisible in single-report analyses. This comparative orientation is consistent with document analysis studies that examine institutional documents across multiple sites or organisations to surface patterns in professional practice \cite{bowen2009document, rapley2007doing}.

\subsection{Document Corpus and Selection Criteria}
\label{sec:document_corpus}

For the purpose of this study, we selected \textit{four representative independent audit reports} assessing three major platform operators and their four distinct VLOPs. 
\begin{itemize}
    \item For Google Ireland Limited, we analysed the DSA audit report for YouTube conducted by Ernst and Young Global Limited (EY), published as the ``2024 Google Ireland Limited DSA Audit Report -- Non-Confidential Version'' \cite{EY2024GoogleDSA}. 
    \item For Meta Platforms Ireland Limited, we examined two separate audit reports: the ``Independent Audit on Facebook'' \cite{ey2024facebookdsa} and the ``Independent Audit on Instagram'' both conducted by EY \cite{ey2024instagramdsa}.
    \item For TikTok Technology Limited, we analysed the ``DSA Assurance Report'' conducted by KPMG, published September 9, 2024, as an ``Independent practitioner's assurance report concerning Regulation (EU) 2022/2065, the Digital Services Act'' \cite{kpmg2024tiktokdsa}.
\end{itemize}

The selected platforms share a common characteristic -- they all employ complex algorithmic systems for content moderation, recommendation, and risk assessment, making them ideal cases for examining how auditors approach technical compliance verification in algorithmically-driven environments.

The resulting analysed corpus consisted of four publicly available audit reports totalling approximately 180 pages of primary documentary material. Document selection followed purposive sampling criteria \cite{patton2015qualitative}: (1) all reports are from the first wave of DSA Article 37 audits published in Q4 2024, ensuring temporal comparability; (2) reports cover designated VLOPs with algorithmic recommender systems, ensuring relevance to our research focus on AI system assessment; (3) reports are publicly accessible non-confidential versions, ensuring reproducibility; and (4) reports cover the same regulatory period (August 2023 to June 2024), enabling direct comparison. These criteria yielded the complete population of published first-wave DSA audit reports from major social media platforms at the time of analysis.

Within the selected documents, the analysis focused on audit methodologies for three critical DSA provisions concerning algorithmic systems. 
\begin{itemize}
    \item The provision on recommender system transparency and non-profiling options under Article 27(3) and Article 38, requiring platforms to provide users with meaningful control over recommender systems. 
    \item The provision on protection of minors under Article 28(2), which prohibits profiling-based advertising to minors when platforms are aware ``with reasonable certainty'' that users are under 18.
    \item The provision on advertisements under Article 26(3), which prohibits advertisement targeting based on profiling using ``special categories of personal data'' as defined in GDPR Article 9(1). 
\end{itemize}
These provisions were purposefully selected based on three criteria: (1) direct relevance to AI-powered algorithmic systems, (2) requirement for technical verification beyond documentary review, and (3) cross-platform comparability of identical legal obligations.

\subsection{Analytical Framework and Coding Procedure}
Our analytical process followed thematic analysis procedures \cite{braun2006thematic} with a hybrid deductive-inductive coding approach \cite{fereday2006rigor}, proceeding through five phases:

\begin{itemize}
    \item \textbf{Phase 1: Familiarisation.} Two authors independently read each audit report in full, generating initial annotations on sections relevant to the three selected DSA provisions. This phase established familiarity with the depth, scope, and structure of each report.
    \item \textbf{Phase 2: Generating initial codes.} We derived coding categories deductively from the DSA's regulatory structure and thus included: (a) stated verification methods (e.g., code inspection, interface testing, document review), (b) evidence types cited (e.g., platform declarations, independent testing), (c) scope limitations acknowledged, and (d) compliance conclusions reached. Additional codes emerged inductively during iterative reading, including: (e) definitional interpretations of ambiguous legal terms (e.g., "reasonable certainty"), (f) temporal constraints in assessment, and (g) methodological variations between auditors assessing identical requirements. Initial codes remained close to the reports' own terminology (e.g., "inspected," "verified," "ascertained"), attending to both manifest content and latent content \cite{graneheim2004qualitative}. Our analysis distinguished between instances where auditors relied on platform declarations versus direct technical verification methods.
    \item \textbf{Phase 3: Constructing themes.} We created a structured extraction matrix with rows for each platform-provision combination (12 cells: 4 platforms × 3 provisions) and columns for the coding categories identified above. Patterns across this matrix were grouped into candidate themes regarding audit methodology characteristics.
    \item \textbf{Phase 4: Reviewing themes.} Candidate themes were reviewed against the coded extracts and the full dataset to ensure internal coherence and adequate evidentiary support. This phase involved collapsing, splitting, and refining themes through discussion between authors.
    \item \textbf{Phase 5: Defining and naming themes}. We defined the scope and boundaries of each theme, producing the challenge categories presented in Section 4.4 (e.g., ``Structural Temporal Assessment Limitation," ``Evaluation Metric Dependency Problem").
\end{itemize}

Throughout this process, we employed constant comparison \cite{glaser1967grounded}, continuously comparing newly coded segments against previously coded material to ensure consistent category application. Where interpretive disagreements arose, we resolved them through discussion until consensus was reached.

\subsection{Researcher Positionality and Limitations}

The examination of audit reports was conducted by authors with complementary expertise in legal analysis (particularly EU digital regulation) and computer science (with specific focus on AI-based systems and algorithmic auditing). This interdisciplinary positioning shaped our analytical lens in a way where we approached the reports asking not only whether auditors fulfilled formal regulatory requirements, but also whether their described methods could plausibly assess the algorithmic behaviours the DSA seeks to regulate. We acknowledge this orientation may foreground technical limitations that auditors operating within established professional frameworks might view differently.
Several limitations may constrain our analysis. First, we analysed only publicly available non-confidential versions of audit reports. Confidential annexes containing detailed technical procedures were inaccessible to wider public and research institutions. Second, we cannot verify whether described audit procedures were actually implemented as stated and our analysis addresses what auditors reported they did rather than what they did. Third, as DSA audits will be conducted periodically each year, these reports will represent an evolving practice and thus subsequent audit cycles may reflect methodological developments not captured in the first wave of reports. Finally, document analysis cannot access auditor reasoning beyond what is explicitly documented; while direct interviews with auditors could complement this analysis, such interviews may be infeasible in practice due to auditors' NDA commitments.

\section{Analysis of Audit Reports}

Results of our examination are structured around three cases corresponding to three critical DSA articles that require auditors to assess algorithmic systems and their compliance mechanisms (as defined in Section \ref{sec:document_corpus}). These cases are then examined collectively in a follow-up section focused on identification of challenges in auditing AI algorithms where we interpret the results, analyse emerging patterns, and discuss implications for regulatory effectiveness and algorithmic governance.

\subsection{Case 1 -- Recommender System Transparency and Non-Profiling Options}

In the first examined case, we focus on transparency and design of recommender systems. Online platform that deploys recommender system shall communicate how these systems operate, including the main parameters used to determine what content is shown to users in their terms and conditions (DSA, Article 27 (1) and (2)). Furthermore, users should be informed about whether content is personalized based on profiling or behavioural data and must be given options to modify or opt out of such personalization (DSA, Article 27 (3) and 38).

For Article 27(3), requiring platforms to provide users with meaningful control over recommender systems, the audits demonstrate varying verification approaches. YouTube's EY auditors claimed to have ``inspected the changes to the recommender system outputs before and after modifying the options and determined that the user's selected options influence the main parameters of the recommender system'' \cite{EY2024GoogleDSA}.

Meta's EY auditors, examining both Facebook and Instagram, employed a different approach, stating they ``inspected the Facebook service and ascertained functionality which allowed the user to select and to modify their preferred recommender system option at any time was available and was directly and easily accessible from the online interface'' \cite{ey2024facebookdsa, ey2024instagramdsa}. Notably, the Facebook and Instagram audit reports are nearly identical in their methodology and conclusions, despite these being distinct platforms with different user bases and engagement patterns.

TikTok's KPMG auditors acknowledged limitations in their assessment approach: ``we were not able to determine, when selecting or modifying these options, that the performance of the recommender systems was appropriately altered throughout the entire Evaluation Period, as we could not obtain sufficient evidence to assess all the changes made to the recommender systems'' \cite{kpmg2024tiktokdsa}.

These audits also overlook how users actually experience recommender controls. Research shows that meaningful impact depends not only on the availability of interface options but on whether users can understand them and perceive changes in recommendations in context \cite{Metaxa2021AuditingAlgorithms,lamenduser}. Audits focused solely on point-in-time output checks therefore may miss broader socio-technical considerations.

\subsection{Case 2 -- Protection of Minors}
Under the DSA, online platforms accessible to minors must adopt measures that are both appropriate and proportionate to ensure a safe digital environment for children and adolescents (DSA, Article 28). This obligation applies when a service is open to minors, specifically directed at them, or when the provider is aware that minors form part of its user base (DSA, Recital 71). The duty to implement protective measures requires platforms to assess the potential impact of their services on young users and to adjust their interfaces, functionalities, and recommender systems accordingly. What qualifies as ``appropriate'' depends on the platform’s nature and content, but the principle of proportionality guides the scope of action, ensuring that responses are tailored to the level of risk \cite{Wilman2024DSACommentary}. To support compliance, the European Commission issued guidelines in 2025 that stress the need to counter AI-driven nudging and manipulative design practices  \cite{EuropeanCommission2025MinorsDSA}.
Article 28(2) complements the provision and prohibits profiling-based advertising to minors when platforms are aware ``with reasonable certainty'' that users are under 18.  This requirement places auditors in the position of assessing AI systems designed to infer user age from behavioural patterns. These are among the most technically complex and privacy-sensitive applications of machine learning on these platforms.

YouTube's auditors claimed they ``validated with reasonable accuracy that it properly categorizes minors and adults through the consideration of various data sources and inputs'' \cite{EY2024GoogleDSA} though no validation methodology or accuracy metrics were provided. 

Meta's auditors defined ``reasonable certainty'' as the stated age of the recipient of the service \cite{ey2024facebookdsa}, effectively relying on self-reported age data. The platform implemented a complete advertisement elimination strategy for users identified as minors.

TikTok received a ``disclaimer of opinion'' for Article 28(1) due to ongoing European Commission proceedings, leaving age determination capabilities unexamined \cite{kpmg2024tiktokdsa}.

The assessment of minors’ protection similarly reflects a narrow technical view. Research consistently show that risks for young users emerge from socio-technical contexts, including misreported ages, limited privacy literacy, and susceptibility to manipulative or persuasive design \cite{devos}. As a result, audit conclusions based primarily on classification methods or platform declarations capture only part of the risk environment that Article 28 is addressing.

\subsection{Case 3 -- Advertisements}
Article 26(3) prohibits advertisement targeting based on profiling using ``special categories of personal data'' as defined in GDPR Article 9(1), including health data, sexual orientation, political opinions, and other sensitive characteristics. 
This obligation applies only where the definition of advertisement according to the Article 3 (r) DSA is fulfilled. This definition covers promotion of messages presented by online platforms against remuneration, i.e., only paid advertisement or paid boosting of content shall be classified as advertisement \cite{Husovec2024PrinciplesDSA}. However, hidden influencer marketing is not covered by this provision.
This requirement presents auditors with the challenge of verifying that complex AI systems do not use prohibited data categories, either explicitly or through inference from non-sensitive data combinations.

YouTube's auditors employed an ``allow list'' approach where ``advertisers do not have the option to select targeting parameters that use special categories of personal data'' \cite{EY2024GoogleDSA}.

With Meta, EY used contractual delegation, with auditors noting that ``the data controller was contractually responsible for the data used in the advertiser's campaign'' \cite{ey2024facebookdsa,ey2024instagramdsa}.

Regarding TikTok, KPMG implemented keyword-based moderation, with auditors testing that the ``moderation system automatically blocks advertisements that use keywords that are not allowed'' \cite{kpmg2024tiktokdsa}.

Audit procedures for sensitive-data advertising also overlook how users interpret ads in practice. Socio-technical audits demonstrates how targeted advertising harms emerge at the intersection of algorithmic inference and user comprehension \cite{Lam2023SociotechnicalAudits}. Technical verification of targeting settings may not fully capture the socio-technical experience of advertising transparency and potential harm.

\subsection{Identification of Challenges in Auditing AI-based Systems}

\textbf{Compliance Conclusion Patterns}.
The systematic analysis of audit reports reveals notable methodological differences when assessing identical DSA requirements. These variations are particularly evident in provisions requiring evaluation of algorithmic behaviour, where auditors demonstrate substantially different technical approaches and conclusions. Differences in patterns how audits are conducted expose fundamental tensions between traditional audit frameworks and the technical realities of assessing algorithmic systems.

TikTok received negative conclusions specifically when auditors acknowledged assessment limitations, while other platforms received positive assessments despite comparable technical verification challenges. YouTube's validation claims lacked supporting statistical evidence, yet received positive conclusions. Meta's definitional approach to ``reasonable certainty'' effectively circumvented technical assessment requirements while achieving positive compliance ratings.

\textbf{Interpreting Audit Methodology Variations}. 
The observed methodological differences raise questions about whether current auditing standards provide sufficient guidance for assessing algorithmic systems. The identical conclusions for Facebook and Instagram suggest audit approaches may lack granularity to distinguish between different algorithmic implementations, while the significant variation in technical depth across auditors indicates uncertainty about what constitutes adequate evidence for AI system compliance.
The pattern where explicit acknowledgment of assessment limitations correlates with negative compliance conclusions, while unsubstantiated claims receive positive ratings, suggests potential inconsistencies in how audit confidence is interpreted. This prompts one to question whether traditional audit frameworks are equipped to handle the unique verification challenges presented by AI systems.

\textbf{Technical Assessment Challenges}. 
The recommender system audits (Articles 27 (3) and 38 (1)) reveal challenges in measuring whether user controls provide meaningful influence over algorithmic behaviour. YouTube's claim to have verified ``influence'' lacks quantitative assessment of control effectiveness magnitude. When users modify their preferences, auditors can verify these actions occurred but cannot measure their actual impact on future algorithmic outputs relative to other system inputs.
Similarly, for non-profiling options under Article 38 (1), auditors can verify that alternative recommendation modes exist but cannot assess whether these truly eliminate profiling or simply reduce certain data inputs while maintaining similar behavioural targeting through other means. This measurement gap suggests current audit approaches may have difficulty distinguishing between superficial compliance controls and genuinely effective user influence mechanisms.

\textbf{The Profiling Classification Challenge}. 
Article 28 (2)'s prohibition on profiling-based advertising to minors presents challenges in determining what constitutes ``profiling'' versus other forms of content selection (e.g., by means of popularity or novelty). To this end, the DSA provides no explicit definition, as well as there is no consensus within the research community on standard metrics or thresholds to determine when a user is considered to be profiled. Meta's complete elimination strategy sidesteps this classification requirement entirely, while YouTube's approach of disabling ``ads personalization'' may not address all forms of behavioural targeting that could constitute profiling under the GDPR definition.
This classification challenge reflects broader difficulties in applying legal definitions of "profiling" to complex advertising systems where personalization occurs through various algorithmic methods that may not clearly fall into profiling versus non-profiling categories.

\textbf{The Advertisement Targeting Verification Challenge}. 
The advertisement targeting audits reveal difficulties in verifying compliance with special category data restrictions under Article 26 (3). TikTok's keyword-based verification can confirm that explicit prohibited terms are blocked but cannot assess whether functionally equivalent targeting occurs through behavioural pattern analysis. YouTube's allow-list approach prevents direct access to prohibited categories but cannot evaluate whether permitted targeting parameters achieve similar audience segmentation.
This verification challenge highlights how compliance assessment focuses on explicit rule implementation rather than functional outcomes, where advertising systems may achieve prohibited targeting effects through technically permitted methods.

\textbf{The Structural Temporal Assessment Limitation}. 
The temporal dimension reveals a structural limitation of traditional audit methodologies when applied to algorithmic systems. Unlike conventional IT systems that remain relatively stable between updates, AI systems continuously evolve through machine learning processes, A/B testing, and algorithmic refinements. Traditional audits, by design, provide point-in-time assessments that cannot capture this dynamic behaviour.
TikTok's auditors explicitly acknowledged this structural constraint, stating they could not determine whether recommender system modifications ``appropriately altered" performance ``throughout the entire Evaluation Period'' as recommender systems continuously retrain on new data. This acknowledgment reveals not merely a procedural limitation but a fundamental inability of traditional audit approaches to assess the sustained effectiveness of algorithmic compliance measures over time.

The rapid pace of platform evolution compounds this temporal limitation. Recent research on algorithmic auditing demonstrates that social media platforms undergo frequent interface changes and algorithmic modifications that render audit findings obsolete within short timeframes \cite{10.1145/3726302.3730293}. This platform instability means that even successful point-in-time assessments may not reflect system behaviour days or weeks later, fundamentally undermining the temporal validity of traditional audit conclusions. The dynamic nature of digital platform features is exemplified by research on TikTok's algorithmic explanations \cite{mousavi2024auditing}, where we documented explanatory features that were accessible during the initial study but had disappeared by the time of writing this paper. This disappearance of auditable features between research phases illustrates how platform modifications can render audit findings obsolete, challenging the assumption that compliance verification remains valid beyond the immediate assessment period.

This temporal limitation is particularly significant because algorithmic systems may exhibit compliance degradation as they adapt to new data patterns, seasonal user behaviours, or trending content. Traditional audit methodologies cannot detect such degradation, creating temporal gaps in compliance verification that are inherent to the methodology rather than correctable through improved procedures.
The structural nature of this temporal limitation distinguishes it from other audit challenges. While measurement precision or classification difficulties might be addressed through better auditor training or standardized procedures, the point-in-time constraint of traditional auditing cannot be overcome while maintaining the fundamental audit approach.

\textbf{The Evaluation Metric Dependency Problem}. 
The audit analysis reveals a fundamental methodological challenge -- compliance conclusions vary significantly based on evaluation metric selection, even when assessing identical systems. This metric dependency problem reflects the difficulty of translating abstract legal requirements into measurable technical criteria, especially in cases when there are no standard widely-accepted measurement procedures or metrics.
Recent research confirms that audit outcomes heavily depend on evaluation metric selection \cite{10.1145/3726302.3730293}, suggesting that the methodological variations observed across DSA audits may reflect different measurement approaches rather than actual compliance differences. This evaluation dependency manifests across the audit evidence in several ways:
\begin{itemize}
    \item For recommender system control effectiveness (Articles 27 (3) and 38 (1)), YouTube's auditors measured ``influence'' through before-and-after output comparison, while Meta's auditors assessed control ``accessibility'' through interface verification. TikTok's auditors attempted longitudinal behaviour assessment but acknowledged measurement limitations. Each approach operationalizes ``meaningful control'' differently, potentially yielding different compliance conclusions for identical user control implementations.
    \item For age determination compliance (Article 28), YouTube's auditors claimed validation through accuracy assessment, while Meta's auditors defined compliance through self-reported age reliance. These represent fundamentally different compliance metrics -- statistical model performance versus data source specification -- applied to the same regulatory requirement of ``reasonable certainty''.
    \item For advertisement targeting restrictions (Article 26 (3)), the audits demonstrate three distinct measurement approaches: keyword-based violation detection (TikTok), targeting option restriction verification (YouTube), and contractual responsibility assessment (Meta). Each metric captures different aspects of compliance while potentially missing violations detectable through alternative measurement frameworks.
\end{itemize}

This evaluation metric dependency suggests that audit outcomes may reflect methodological choices rather than actual platform compliance differences. The absence of standardized measurement criteria for algorithmic compliance means that auditors must translate legal requirements into technical metrics using professional judgment, introducing systematic variability that undermines regulatory consistency. Unlike traditional IT audits where compliance metrics are often well-established, algorithmic system assessment requires novel measurement approaches that are not yet standardized across the auditing profession.

\textbf{Missing Socio-technical Dimensions}. The absence of socio-technical framing, which is understood as the recognition that algorithmic systems operate as situated collections of technical components, user practices and design choices \cite{Lam2023SociotechnicalAudits, leerssen2020soap, Metaxa2021AuditingAlgorithms}, repetitively emerged from analysed reports. These audits conceptualize platforms solely as technical artefacts whose compliance can be verified through point-in-time checks of configuration settings, interviews with platform's governance officers, documentation, or static interface elements. However, recommender systems and ad-delivery algorithms are socio-technical infrastructures meaning that their behaviour is shaped not only by code and data pipelines but by user practices, interface design and emergent feedback loops between human activity and algorithmic adaptation. Without acknowledging these aspects, audits risk producing findings that are technically accurate yet substantively incomplete.

\begin{table*}[t]
\centering
\caption{Complementarity of traditional and algorithmic auditing approaches.}
\label{tab:audit-comparison}
\footnotesize
\renewcommand{\arraystretch}{1.4}

\begin{tabular}{>{\raggedright\arraybackslash}p{2.5cm} >{\raggedright\arraybackslash}p{5.5cm} >{\raggedright\arraybackslash}p{5.5cm}}
\toprule
 & \textbf{Traditional Auditing} & \textbf{Algorithmic Auditing} \\
\midrule

\textbf{Time-span coverage} & 
Point-in-time assessment, auditors verified functionalities ``were visible and operational'' but acknowledged inability to assess system behaviour throughout the evaluation period & 
Continuous observation and tracking whether control modifications maintain influence over time as users continue interacting and algorithms relearn from new behaviour  \\
\midrule

\textbf{Evidence basis} & 
Documentation review, code inspection, management inquiries; verifies what systems are \textit{designed} to do & 
Simulated user interactions with live systems, observes what algorithms \textit{actually do} in practice \\
\midrule

\textbf{Socio-technical perspective of auditing methodologies} & 
Limited: Platforms conceptualized as technical artefacts; compliance verified through configuration checks, documentation, and interface inspection; risks producing findings that are technically accurate yet substantively incomplete & Integrated Approach: When designed with attention to user interactions, it can capture how algorithmic behaviour emerges from user practices, interface design, and feedback loops between human activity and algorithmic adaptation \\
\midrule

\textbf{Compliance criteria} & 
Auditors independently translate legal requirements into metrics; different operationalisations yield incomparable conclusions across platforms & 
Structured evaluation with predefined metrics; same measurement approach enables cross-platform comparison \\
\midrule

\textbf{Independence} & 
Limited: conducted by auditing companies selected and paid by platforms, relying on platform-provided documentation and benchmarks & 
High: fully external, available for researchers, civil society and serving as a powerful watchdog tool that is platform-agnostic and representative \\

\midrule

\textbf{Scope} &
Broad: can assess full range of DSA obligations including organizational measures, security practices, content moderation policies, transparency reporting, and governance structures &
Narrow: limited to behavioural assessment of algorithmic systems, such as recommender systems and ad targeting; cannot evaluate organizational compliance, security measures, or policy documentation\\

\bottomrule
\end{tabular}
\end{table*}

The identified challenges point toward the need for assessment methodologies specifically designed for dynamic algorithmic systems. While various approaches could strengthen oversight, such as including enhanced user controls or standardized audit criteria, \textit{we advocate for algorithmic auditing as a complementary methodology capable of addressing the structural limitations of conventional approaches (we summarise the complementarity of these paradigms in Table~\ref{tab:audit-comparison})}.

Traditional audits remain essential for assessing the full breadth of DSA obligations, including organizational measures, security practices, and governance structures, that fall outside the scope of behavioural algorithm assessment. Algorithmic auditing should not replace this broader compliance framework but strengthen it where conventional methods face inherent limitations: in capturing the dynamic, adaptive behaviour of AI systems over time. Section 5 develops this proposal in greater detail.

\section{Social Media Algorithmic Auditing}
The challenges of traditional audit methodologies, identified through the previous works (Section 2) as well as our own audit analysis (Section 4), suggest that such auditing approaches face structural limitations when applied to dynamic and complex algorithmic systems. The temporal constraint, in particular, represents a categorical inadequacy rather than a procedural variation, as traditional audit methodologies are inherently designed for static system assessment. In addition, applied methodologies remain limited to mostly formal assessment of implementation of DSA requirements (e.g., when assessing an ability of users to control recommender systems, some audits focused purely on presence of such options in user interface). We stress that it is necessary to go beyond such single-shot, formal and static assessment and investigate AI systems dynamically -- i.e., how they interact and respond to users with various characteristics reflecting the ``spirit'' of the DSA.

To achieve this, we argue for more prominent use of \textit{algorithmic auditing}. This paradigm, by contrast to compliance-oriented audits, has roots in social science audit work dating back to the 1960s \cite{Metaxa2021AuditingAlgorithms} and has been adapted as a methodology for investigating algorithmic systems  ``from the outside in'', which allows for systematically querying systems with controlled inputs to draw inferences about opaque behaviours \cite{Sandvig2014Audits,Metaxa2021AuditingAlgorithms}. Algorithmic audits can be defined as ``assessments of the algorithm’s negative impact on the rights and interests of stakeholders, with a corresponding identification of situations and/or features of the algorithm that give rise to these negative impacts''~\cite{10.1177/2053951720983865}. Since social media AI algorithms are black-boxes (we cannot analyse or influence their inner workings), audits must explore their properties behaviourally: user interactions with an AI algorithm are simulated (e.g., content visits), and observed responses (e.g., recommended items) are examined for the presence of the audited phenomenon. Typically, bots~\cite{10.1145/3392854,yaudit-recsys} or human agents~\cite{10.1145/3351095.3372879} are employed to simulate such user interactions in so-called sockpuppeting audits and crowdsourcing (collaborative) audits respectively~\cite{Sandvig2014Audits}. 

\begin{figure}[t]
  \centering
  \includegraphics[width=0.95\linewidth]{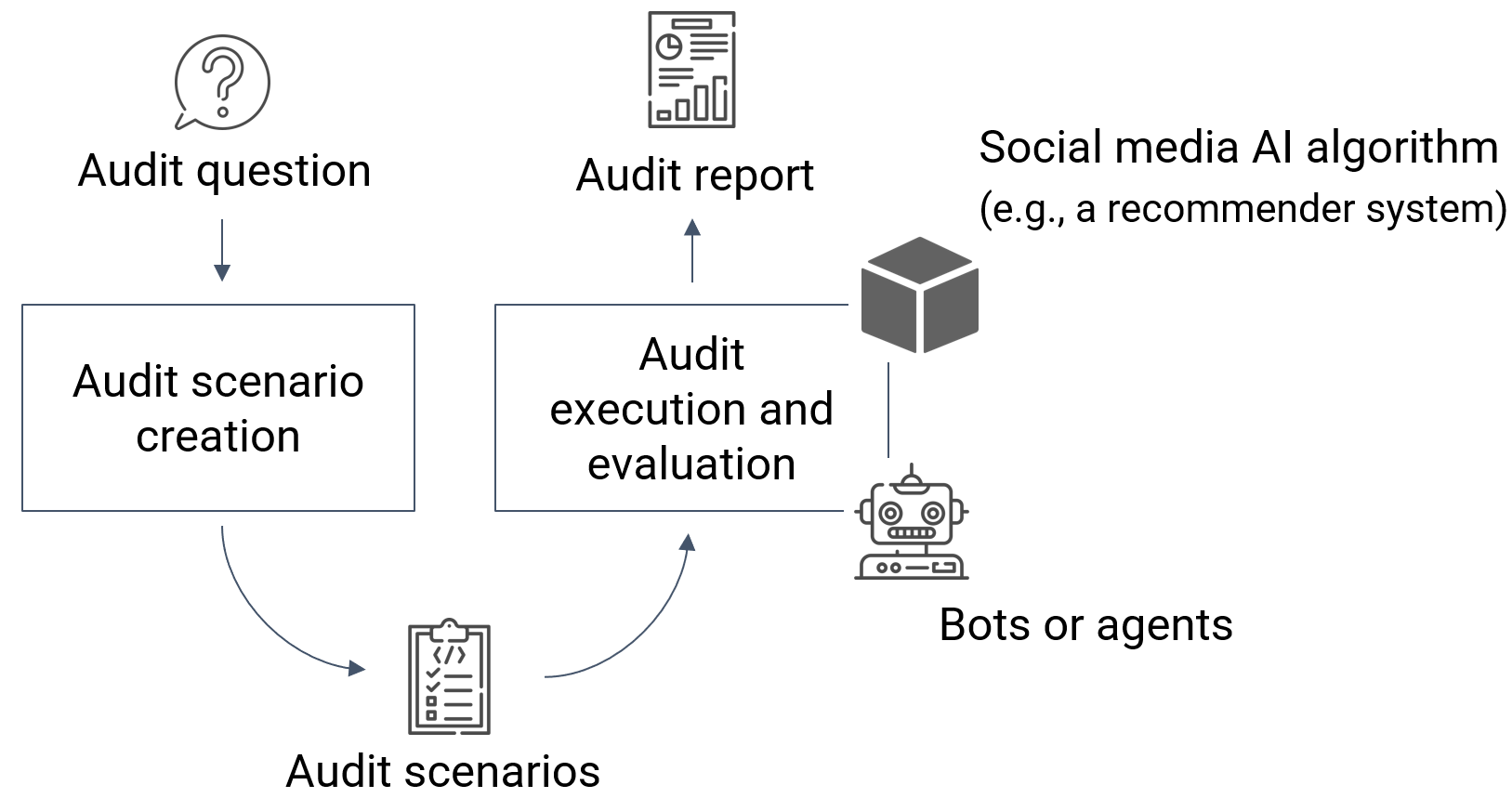}
  \caption{A process of typical algorithmic auditing approach. It consists of the following steps: 1) by proceeding from an audit question, audit scenarios consisting of user profiles and user actions are created; 2) during audit execution and evaluation, bots/human agents execute such audit scenarios, platform responses are recorded, and further investigated for the presence of audited phenomena; 3) audit reports are produced.}
  \Description{A typical algorithmic audit consists of the following steps: 1) by proceeding from an audit question, so called audit scenarios consisting of user profiles and user actions that should be used to simulate user behaviour are created; 2) during audit execution and evaluation, bots/human agents execute the created audit scenarios, platform responses are recorded, and further investigated for the presence of audited phenomena; 3) audit reports are produced.}
  \label{fig:algorithmic-auditing-process}
\end{figure}

A typical algorithmic audit consists of the following steps (Figure \ref{fig:algorithmic-auditing-process}). First, by proceeding from an \textit{audit question}, so called \textit{audit scenarios} are specified. Audit scenarios consist of user profiles (personal information, user’s interaction history) and user actions that should be used to simulate user behaviour (interactions with the pre-selected or presented/recommended content, queries to search, etc.). Next, during \textit{audit execution and evaluation}, bots/human agents execute the created audit scenarios, platform responses are recorded, and further investigated for the presence of audited phenomena (e.g., a presence of harmful content or a level of personalization). Finally, \textit{audit reports} are produced. The current algorithmic audits, especially due to their recent emergence, are performed mostly by researchers, but we can already also observe successful adoption by practitioners, NGOs or media (e.g., Wall Street Journal investigation\footnote{\url{https://www.wsj.com/tech/tiktok-algorithm-video-investigation-11626877477}} of TikTok recommender system conducted in July 2021). 

While algorithmic auditing provides a promising behavioural approach, its effectiveness depends on embedding human-centered principles into the design of audit scenarios. Algorithmic auditing can also be understood as a methodology in line with HCI visions. Although typically described as a technical behavioural assessment, its core practices including designing synthetic users, modelling interaction patterns, and evaluating system responses reflect the central HCI principle of studying systems through the lens of human (user) use. Sockpuppet profiles and simulated interaction trajectories function as proxies for real individuals navigating platforms in real conditions, enabling auditors to reproduce, at scale, the types of behaviours, routines, and interpretive strategies that shape algorithmic outputs in everyday life. This moves auditing beyond abstract system inspection and situates it squarely within the socio-technical environment where risks actually emerge.

An important feature of algorithmic audits is that they are, from their inherent design, external and independent from social media platform being audited. Therefore, external entities, including researchers, may play a crucial oversight role in the auditing process. They may contribute in three ways:
\begin{enumerate}
    \item Conducting independent audits as recognised auditing bodies.
    \item Verifying audit results to ensure they are accurate and unbiased.
    \item Analysing audit credibility by reviewing whether platforms are making genuine compliance efforts or just aiming for regulatory approval.
\end{enumerate}

\subsection{Relevance to Identified Challenges in Auditing AI-based Systems}
Algorithmic audits have a potential to address multiple of the identified challenges presented in Section 4.4.

First to address the identified \textit{Structural Temporal Assessment Limitation}, the dynamic nature of the social media platforms, where the behaviour is constantly adapted and tailored to specific users based on learning their behaviour, necessitates continuous long-term and large-scale assessment (e.g., days or weeks of interaction on the platform) across different users \cite{simko2021continuous-automatic-audits}. Since algorithmic audit allows an appropriate level of automation (i.e., simulating user behaviour, data collection, content annotation), they would allow for a long-term assessment without requiring significant manual labour. The audits can be executed over long period of time (weeks/months), allowing to deal with the constantly changing platform behaviour. As such, this would also allow the auditors to determine, whether the implemented system modification has led to better compliance. Furthermore, it would also allow to determine whether the compliance is implemented only for a specific time period (or specific users) or whether it is done long-term.

In addition, using such automation would allow for assessing the different obligations more directly by tailoring the synthetic users, their behaviour and the setup specifically to the assessment (addressing e.g., \textit{the Advertisement Targeting Verification Challenge}). For example, if we want to assess the protection of minors obligations, comparing how the social media platform recommendation behaves when users of different age groups, but same interests, engage with the platforms. The same can be done for other obligations as well, e.g., observing whether selecting the non-profiling option changes anything or whether advertisements are done based on protected characteristics.

Finally, while the algorithmic audits will not fully solve some of identified challenges (e.g., \textit{the Profiling Classification Challenge} or \textit{the Evaluation Metric Dependency Problem}), they implement a specific structured evaluation -- by applying specific methodologies how the assessment is done and what metric is used. Such well-defined structured methodologies allow easier reproducibility, standardization and applicability across different social media platforms. As such, it minimizes the potential differences in the audit findings due to differences in methodology. 

While current generation of algorithmic audits represent a promising approach to assess compliance with the DSA obligations, a deeper insight still shows presence of drawbacks hampering such larger adoption in practice. First and foremost, implementation of algorithmic audits is technically demanding. Algorithmic audits require to implement specific agents for each platform being audited, as well as their adjustments for each specific audit question. Second, current audits still remain quite oversimplified and artificial, and generally do not reflect the complexity of real-world social media environment. For example, audit scenarios, which are created entirely manually and ad-hoc on intuition of researchers, remain incomplete (audits cover only a small subset of relevant content/user/interaction space) and inauthentic (simulated behaviour is heavily prescribed and does not correspond to how real users actually behave). However, this issue can be (at least partially) solved by the recently proposed novel algorithmic paradigms, such as model-based algorithmic auditing \cite{srba2025model}. Finally, the recent study showed that current audits suffer with low replicability \cite{10.1145/3726302.3730293}, which is crucial when such audits should be applied in practice. While these issues may negatively effect the adoption of algorithmic audits in enforcement of legal obligations, we may expect that the situation will improve thanks to the significant research and practitioners' efforts in this area.

\subsection{Illustrative Use-Case for Auditing DSA Obligations}

To illustrate how algorithmic audits can be utilized to audit DSA obligations, we choose the profiling of minors (Article 28) as a use-case, directly addressing the identified \textit{Advertisement Targeting Verification Challenge}. The audit question in this case is whether the social media recommender system suggests a lower number of personalized advertisements for minors compared to adult users. For audit scenarios, we can prepare 2 separate sets of users that differ only in age: 1) minors, e.g., users in the 14-17 age range; and 2) adults, e.g., users in the 18-25 age range. For these users we specify topics of interests (e.g., beauty, gaming, fitness) and uniformly distribute them across these topics and other user characteristics, such as gender and location in one of the EU countries. To build the watch history, we begin by searching for and watching videos that belong to the topic of interest. Afterwards, during the audit execution, the users connect to the social media platform and spend some time there, interacting with the content. How often the users connect on each day, for how long they interact and how long the audit runs are all defined by the audit scenario. For example, the audit can run for 20 days, where each day the user will interact with the content for up to one hour. As it is all automatic and can be run long-term or repetitively, this directly addresses another identified challenge of the \textit{Structural Temporal Assessment Limitation}. After the study is finished, both sets are evaluated based on how many personalized advertisements they observed during the audit, i.e., advertisements related to their topic of interest. Based on the number of such advertisements observed (or the ratio of relevant and non-relevant ads), we can determine whether minors are being profiled for advertisement -- if the amount of personalized ads is not statistically significantly lower compared with adult users, we can conclude that the profiling is happening. By using the same metric, the audit also addresses the \textit{Evaluation Metric Dependency Problem}. Based on the results from the social media platform (or multiple platforms), we can prepare the audit report that highlights the identified issues and compares between platforms. A similar study can also be run for additional DSA obligations, simply by changing the user characteristics and the scenario, e.g., instead of minors, using a set of users with personal characteristics that should not be used for personalization (Article 26 (3)).

\section{Conclusions}

Our systematic analysis of audit methodologies across three critical DSA provisions related to AI algorithms used by social media reveals a consistent pattern: traditional audit approaches encounter significant challenges when applied to AI systems whose complexity and dynamism exceed conventional assessment capabilities. Auditors are tasked with verifying algorithmic behaviours that may not be fully observable through code inspection, management interviews, and interface testing -- the standard tools of IT auditing.
The challenges are particularly acute for provisions requiring assessment of algorithmic behaviour rather than system features. Verifying that recommendation controls provide ``meaningful'' user influence, that age inference models operate with ``reasonable certainty'', and that advertising systems avoid prohibited profiling requires understanding of AI system behaviours at a level of detail that may exceed current audit methodologies.
TikTok's auditors demonstrate greater acknowledgment of these limitations, explicitly stating they could not obtain sufficient evidence to assess recommender system behaviour across the entire evaluation period. This honesty about methodological constraints raises questions about how other auditors reached more confident conclusions about similar technical challenges.

The variation in audit approaches across platforms indicates that the profession may still be developing appropriate standards for AI system assessment. The identical conclusions for Facebook and Instagram despite their different characteristics, the dramatically different technical depth between YouTube and Meta audits, and the varying interpretations of legal standards like ``reasonable certainty'' suggest that audit outcomes may reflect methodological choices rather than compliance differences.
This inconsistency undermines the regulatory framework's goal of establishing uniform compliance standards across platforms. If different audit approaches can yield positive compliance conclusions for identical legal requirements, this suggests either that the requirements themselves lack sufficient specificity or that audit methodologies require greater standardization.

The challenges identified in current audit methodologies point toward the need for new approaches specifically designed for AI system assessment. To this end, we advocate for adoption of algorithmic audits -- a process of behavioural assessment of AI algorithms by means of simulating user behaviour, observing algorithm responses and analysing them for audited phenomena. As such algorithmic audits allow automatization of individual auditing steps (e.g., simulation of user behaviour, data annotation), they allow to conduct systematic long-term and large-scale audits (in terms of assessed user characteristics, such as age, gender, location). Despite this great potential, their utilization in auditing processes still requires to overcome multiple technological as well as research challenges, such as how to design audit scenarios to sufficiently cover a great user-content-interaction space present on social media platforms or how to authentically simulate user behaviour. A research and innovation activities addressing these challenges therefore represent a potential for multidisciplinary research, involving experts with computer-science (including AI and recommender system) as well as social-science (including legal and ethics) background.

Ultimately, meaningful DSA compliance requires acknowledging that algorithmic systems operate within socio-technical environments shaped by real user behaviour. Because algorithmic auditing embeds modelled users into live platform conditions, it reveals how algorithms and people together produce outcomes. This aspect traditional audits are not able to capture. Incorporating such human-centred auditing approaches is therefore essential for enforcement mechanisms that reflect how platform algorithms affect users in practice.

\begin{acks}
This work was supported by EU NextGenerationEU through the Recovery and Resilience Plan for Slovakia under the projects No. 09I03-03-V03-00020 and 09I01-03-V04-00100; and by European Union, under the project lorAI - Low Resource Artificial Intelligence, GA No. \href{https://doi.org/10.3030/101136646}{101136646}.
\end{acks}

\bibliographystyle{ACM-Reference-Format}
\bibliography{bibliography}

%\appendix
%\section{Appendix}

\end{document}